\documentclass[11pt]{article}
\usepackage{fullpage,amssymb,clrscode}

\usepackage{amssymb}
\usepackage{amsmath}
\usepackage{subfig}
\usepackage{graphicx}

\usepackage{url}
\urldef{\mailsa}\path|chaoxu3@illinois.edu, skiena@cs.sunysb.edu|

\begin{document}
\bibliographystyle{plain}

\title{Marking Streets to Improve Parking Density}

\author{Chao Xu\thanks{chaoxu3@illinois.edu, Department of Computer Science, University of Illinois at Urbana-Champaign. Supported in part by NSF REU supplement IIS-1128741.} \and Steven Skiena\thanks{Corresponding author. 
skiena@cs.sunysb.edu, Department of Computer Science, Stony Brook University. Supported in part by NSF grants IIS-1017181 and DBI-1060572.
}}

\maketitle

\begin{abstract}
Street parking spots for automobiles are a scarce commodity in most urban environments.
The heterogeneity of car sizes makes it inefficient to rigidly define fixed-sized spots.
Instead, unmarked streets in cities like New York leave placement decisions to individual drivers,
who have no direct incentive to maximize street utilization.

In this paper, we explore the effectiveness of two different
behavioral interventions designed to encourage better parking, namely
(1) educational campaigns to encourage drivers to ``kiss the bumper''
and reduce the distance between themselves and their neighbors, or (2)
painting appropriately-spaced markings on the street and urging drivers to
``hit the line''.
Through analysis and simulation,
we establish that the greatest densities are achieved when lines are painted to create spots
roughly twice the length of average-sized cars.
Kiss-the-bumper campaigns are in principle more effective than hit-the-line for equal degrees of compliance,
although we believe that the visual cues of painted lines induce better parking behavior.
\end{abstract}

\section{Introduction}
Alternate side of the street parking is a unique but important New York City institution.
Most city streets are assigned two intervals per week
(typically 1.5 hours each) during which a particular side of the street must be
vacated to allow for street cleaning.
Drivers double park on the other side of the street
during this time window, 
waiting for the moment when the street cleaner passes or the period expires, at
which point they must quickly move their cars to newly clean and once-again
legal spots.
Alternate side of the street parking defines the rhythm of life
for many city residents \cite{Tepper} by mandating regular
actions in order to maintain a car in the city.

The effective number of New York City parking spots depend heavily
on the dynamics
of alternate side of the street parking, since all cars park
simultaneously but generally depart at distinct times.  Thus it
is rare for two adjacent spots to open simultaneously after the
street configuration is frozen at the end of the forbidden interval.
This means that any extra space left between the hastily-parked cars cannot
be reclaimed until all the vehicles move again during the next parking
switch.   New York streets do not contain any laws, painted lines or
boundary markings to guide positioning or restrict the space
between neighboring cars after this transition.   Since it is usually
easier for the driver to leave ample room between cars, much of the potential
parking area is wasted.

The upshot is that a classical random process fairly accurately
captures parking behavior in the city, and motivates the challenge of
identifying behavioral perturbations which significantly improve it.
In this paper, we study the efficacy of two possible perturbations:
(1) educational campaigns that encourage drivers to ``kiss the bumper''
and reduce the distance between themselves and their neighbors, or (2)
painting appropriately-spaced markings on the street and urging drivers to
``hit the line''.

\begin{figure}
\centering
   \includegraphics[scale=0.95]{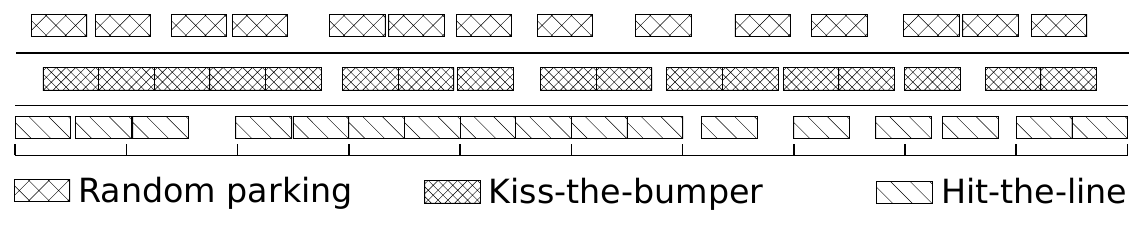}
\caption{Representative street landscapes for random/R\'{e}nyi (top),
kiss-the-bumper (center), and hit-the-line (bottom) for a street of length
$l=20$ with the optimal line spacing of $k=2$,
for $\alpha$ values of 0, $0.5$, and $0.5$ respectively.
\label{parking-cartoon}}
\end{figure}

In this paper, we demonstrate that both mechanisms lead to increasing
density as greater factions of the population employ them.
Our contributions are:

\begin{itemize}

\item
{\em Optimizing Line Spacing} --
Through analysis and simulation,
we establish that the greatest densities are achieved when lines are painted to create spots
roughly twice the length of average-sized cars.

\item
{\em Relative Effectiveness of Interventions} --
Kiss-the-bumper campaigns are in principle more effective than hit-the-line for equal degrees of compliance,
although we believe that the visual cues of painted lines will induce
greater driver compliance.

\end{itemize}

We believe that our results have genuine implications for improving the parking density of New York streets,
and have begun preliminary efforts to interest city transit officials in a pilot study to test these ideas.
More details appear in the conclusions (Section \ref{conclusions}) of this paper.

\section{Related Work}
We have not identified any substantial literature analyzing
the mechanisms or dynamics of
{\em urban} street parking.
Municipal codes typically specify the required dimensions of on-street parking, but do not describe the
rationale behind such a selection.
Representative is 
the City of Bowling Green, Kentucky \cite{bowlinggreen},
which enforces an on-street parking layout that divide
the street into even parking slots of size 6.7m to 7.9m, designed to leave
at least 2.4m of maneuver space.
Different cities have different regulations for on-street parking,
although all must
meet the American Disability Act (ADA) standards for accessibility \cite{ADA-10}.
Design guides to building multistory parking garage structures include \cite{NPA-10,CSBIM-01}
Other studies seek to optimize the revenue from parking facilities \cite{Rojas-06}.

Perhaps the most related paper in terms of our research is Cassady and Korbza \cite{CK-98}, who study the
performance of two different driver search strategies in minimizing (1) walking time, (2) driving time, and (3) combined time
in reaching their destination in a typical parking lot layout.
Like us, they are interested in analyzing the impact of drivers parking strategies.
However, we are concerned with maximizing the utilization of scarce street parking 
in an urban setting.


There has been considerable theoretical work on one particular model of street parking.
The R\'{e}nyi parking problem \cite{Renyi-58} concerns the following random process.
Unit-length cars select unoccupied positions on the real line uniformly at random, starting from an initially empty
interval of length $l$.
The process continues until no vacant unit-length interval remains.

Each car parked in an open interval splits the interval into two smaller ones,
each independent from the other.
Therefore this random process can be analyzed by a recursive formulae.
Let $f(l)$ denote the expected number of cars parked on a street of length $l$.
Then $f(l)$ is determined by the following delay differential equation:

\begin{equation}
f(l) = 1+\frac{1}{l-1} \int_0^{l-1} (f(x) + f(l-x-1))\; dx= 1+\frac{2}{l-1} \int_0^{l-1} f(x) \; dx
\end{equation}
with the base cases $f(l) = 0$ for $l\in[0,1)$ \cite{onparking}.

The parking density is defined as $f(l)/l$, with the R\'{e}nyi parking constant giving the limit of density
as $l \to \infty$.
Numerical calculations show that
\[\lim_{l \to \infty} f(l)/l \approx 0.7476\]
For small integral values of $l$, exact values of $f(l)$ are known, including
$f(1) = 1$, $f(2) = 1/2$, $f(3) = 2/3$, and $f(4) = (11-4 \ln{2})/3$.
The variance in this process has been determined \cite{onparking,Mannion-64}. 
Recently, the expected number of gaps of certain size was analyzed for a discrete version of the parking problem \cite{ClaySimanyi-14}. 

The problem has been generalized into higher dimensions.
Here the goal is to  ``park'' unit $n$-cubes into a larger $n$-dimensional cube.
Results in two dimensions include \cite{BS-70,Palasti-60}.
Higher-dimensional versions are widely studied in statistical
physics as the random sequential adsorption problem. It is investigated as the end configuration
after molecules attaching itself to a surface. See \cite{cmp} for a survey of the field.
The limit of the parking density in $n$-dimensions is a difficult problem, and even the two-dimensional version
remains open \cite{finch}. 

These investigations suggest a new class of bin packing or knapsack
problems \cite{KPP-04,LMM-02}
where the heuristic employed are not under central control,
but instead the heuristic is selected at random to position
each additional element.

\section{Models of Parking Behavior}

Painted street markings are the traditional way to enforce parking behavior.
Each slot must accommodate the largest-sized vehicle on
the road (the GMC Savana van, 224 inches long), leaving considerable wasted space when
small vehicles (e.g. the Nissan Cube, 158 inches) are parked.
The longest cars on the road
approach twice the length of the shortest,
making maximally-generous spots too wasteful for the city to enforce.

Instead, we consider lines as {\em optional} guide markings to help drivers make
better decisions on where to place their car.
These lines serve the same role as etching images of flies in men's urinals:
providing a target to hit that encourages better behavior.
Schiphol Airport in Amsterdam reports a 80\% reduction in spillage as a result
of this design \cite{Routledge-04}.

In this paper, we will study three basic parking strategies 
(illustrated in Figure \ref{parking-cartoon})
where observing street markings is optional:

\begin{itemize}
\item {\em Random parking} --
Consistent with the R\'{e}nyi model, the driver selects a location
uniformly at random over all unoccupied locations on the given street.
We believe this is a fairly accurate model of what happens
in practice during the phase transition from double parking at the 
instant the other side of the street becomes legal again.
\item {\em Kiss the bumper} --
Here a driver selects a location uniformly at random and moves up to leave minimum space
between the neighboring car.
Uniformly observed, this strategy provides optimal street utilization, but 
the parallel maneuvering during the shift
makes it impossible to enforce such behavior.
\item {\em Hit the line} --
Here a driver selects an unoccupied street marking (line) uniformly at random and lines up immediately
behind it.
In the event that all lines are occupied, the kiss the bumper strategy is
employed.
\end{itemize}

We consider the situation when
the population employs a pair of these basic strategies, typically the
random/R\'{e}nyi strategy and a more sophisticated algorithm.
Let $\alpha$ denote that fraction of the population employing the more
sophisticated strategy, i.e. the ``good'' drivers.
The expected density of cars parked on the street depends upon $\alpha$.
For $\alpha=0$, all drivers are random/R\'{e}nyi drivers, so the achievable
density will be $0.7476$.
For $\alpha=1$, all drivers behave in a socially-minded way, and the resulting
density approaches the optimal value of $1.0$.

The R\'{e}nyi recurrence relation can be readily generalized to deal with a
mix of strategies.
Let $f(l)$ denote the expected number of cars that will be parked on a street of
length $l$ where the $\alpha$-fraction of good drivers employ the
kiss-the-bumper strategy.
Then
\begin{equation}
f(l) = 1 + \alpha f(l-1) + (1-\alpha) \frac{2}{l-1} \int_0^{l-1} f(x) dx
\end{equation}

Alternately, let $f_k(l,t)$ denote the expected number of cars that will be
parked on a street of length $l$ where an $\alpha$-fraction of good drivers
employ hit-the-line.
We assume the street is painted with lines $k$ units apart, such the initial
line is $t$ units away from the origin.
Then
\begin{equation}
f_k(l,t) = 1+\alpha f(l-1) + (1-\alpha) \frac{1}{l-1} \int_0^{l-1} f_k(x,\min(x,t)) + g_k(l,x,t) dx
\end{equation}

\begin{equation}
g_k(l,x,t) = \begin{cases}
f_k(l-x-1,t-x-1)   &\mbox{ if } x+1\leq t\\
f_k(l-x-1,k-((x+1-t)-k\lfloor \frac{x+1-t}{k} \rfloor))   &\mbox{ if } t<x+1\leq t+k\lfloor \frac{l-t}{k}\rfloor\\
f_k(l-x-1,l-x-1)   &\mbox{ if } t+k\lfloor \frac{l-t}{k}\rfloor<x+1\\
\end{cases}
\end{equation}

\begin{equation}
f_k(l,t) = f(l) = 0 \mbox{ if } 0\leq l <1
\end{equation}

These equations are difficult to solve both numerically and analytically.
Therefore we will employ discrete-event simulations to study to performance
of particular strategy mixes.

\section{Line Spacing for Unit-Length Cars}

We now consider the problem of spacing painted lines 
so as to induce the greatest expected parking density
as a function of $\alpha$.
We assume that street markings occur regularly every $k$ units over the
length $l$ street.
For unit-length cars, and compliance fraction $\alpha=1$, clearly $k=1$.
However, we will show that this is not the case when there is lower compliance,
and the $(1-\alpha)$ fraction of ``bad'' drivers employ the R\'{e}nyi strategy.

In particular, through simulation we compare the expected density as a
function of $k$ and $\alpha$.
We also compare it to the benchmark kiss the bumper strategy for the same
value of $\alpha$.
For each strategy, the simulations were run on $l=20$ with 100,000
repetitions on 200 evenly spaced data points of $\alpha$ from 0 to 1. Each data point for a specific $\alpha$ represent the average of the parking density over all repetitions. 

\begin{figure}
\centering
   \includegraphics[scale=0.75]{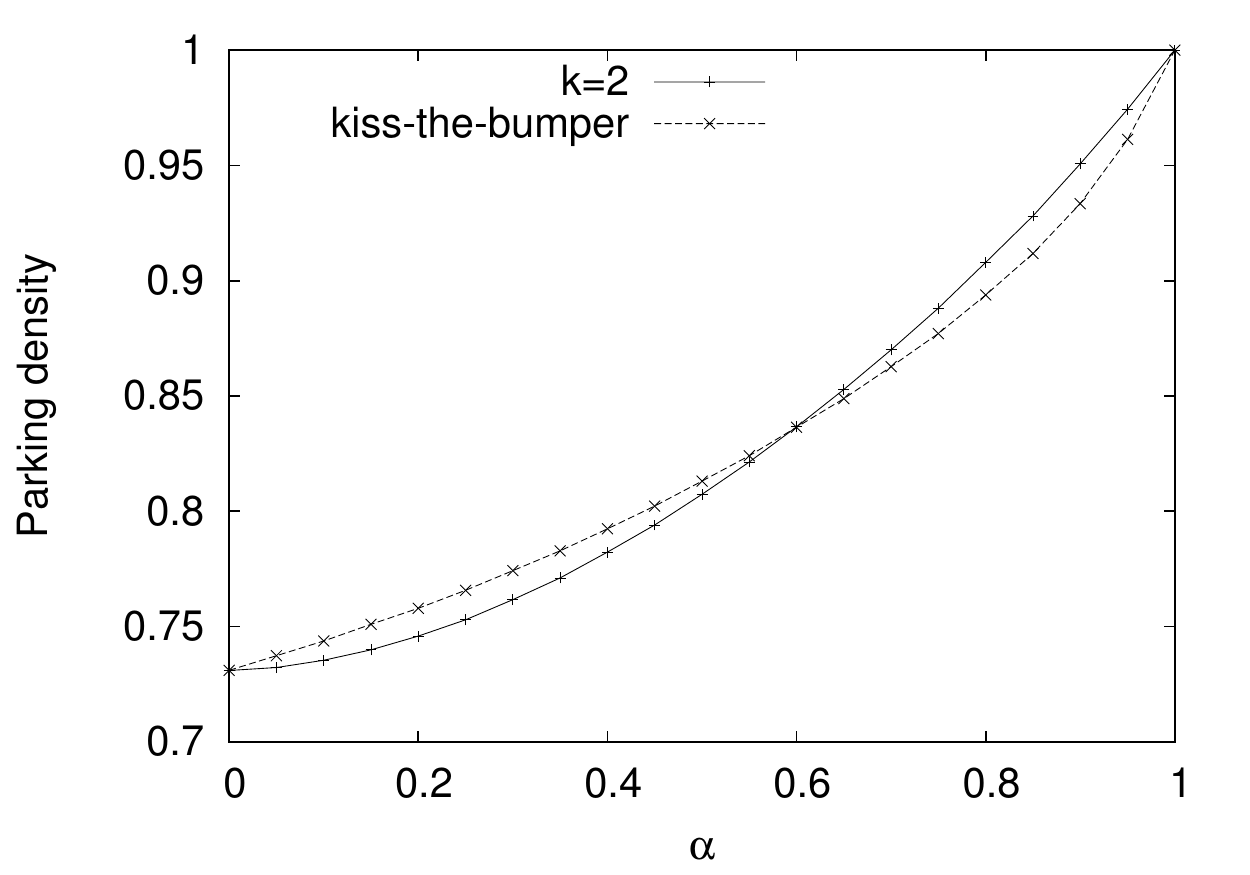}
\caption{Parking density of kiss-the-bumper and hit-the-line as a function
of $\alpha$, for the optimal line spacing of $k=2$ and $l=20$.
\label{main-result-figure}}
\end{figure}

Our primary result is shown in Figure \ref{main-result-figure}.
First observe that the parking density of
both the kiss-the-bumper and hit-the-line strategies range from the R\'{e}nyi
parking constant to an optimal packing as the compliance constant $\alpha$
ranges from zero to one.
The density increases in a non-linear fashion for both strategies, with
densities of 0.82 achieved by $\alpha=0.5$.
This is only about 25\% of the total density improvement with $\alpha=1$, so
relatively high compliance rates must be achieved to substantially
increased density.

The behavior of the two strategies are similar but not identical.
Kiss-the-bumper outperforms hit-the-line until a compliance threshold
of approximately $\alpha=0.592$, beyond which hit-the-line is better.
Here, we employ the optimal line spacing and 
assume that ``good'' drivers kiss-the-bumper when no line is available to hit.

\begin{figure}
\centering
  \subfloat{\includegraphics[scale=0.5]{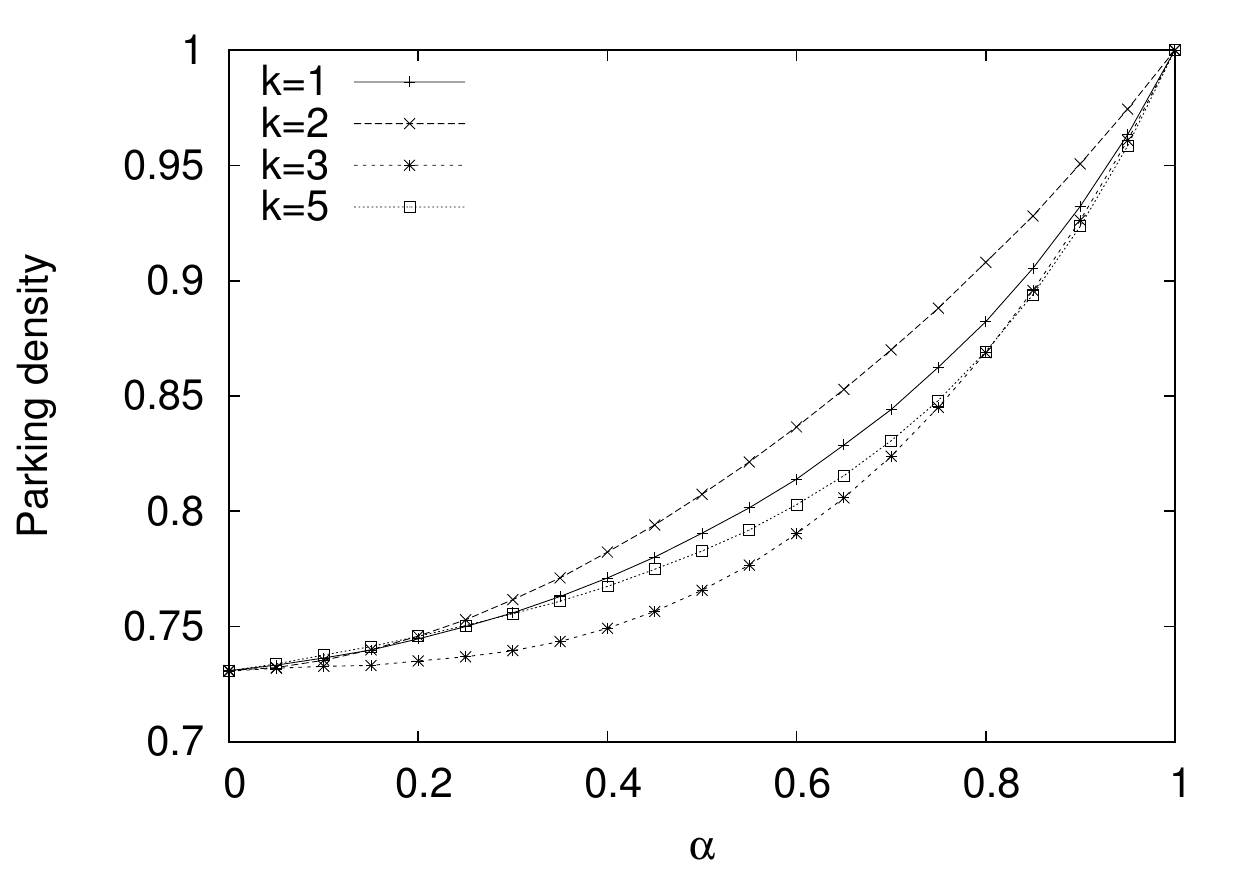}}
  \subfloat{\includegraphics[scale=0.5]{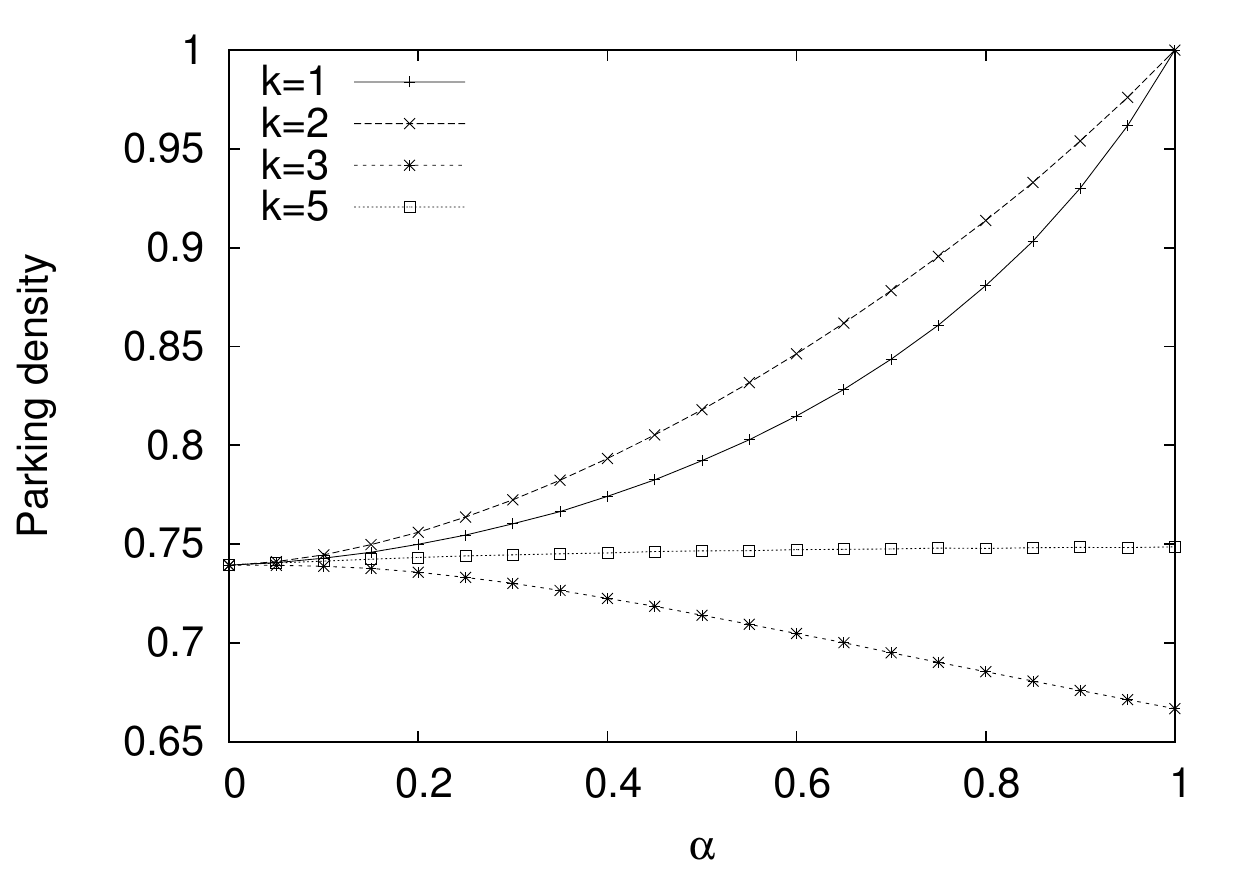}
  \label{fig:density_vs_size:example}}
  \caption{Effect of secondary strategies as a function of line separation:
kiss the bumper (left) vs. R\'{e}nyi random selection (right) for line separations of 1, 2, 3, and 5 car lengths.}
  \label{fig:density_vs_size}
\end{figure}

Figure \ref{fig:density_vs_size}(left) explores parking density as a function
of $k$, the number of car widths between painted lines.
Surprisingly, the best performance is achieved when painting lines between
every {\em other} spot, (i.e. $k=2$), as opposed to the standard practice
of delimiting the precise boundary of every car.
The reason becomes clear in hindsight.
Two ``good'' drivers parked on consecutive lines for $k=2$ will create a
unit-length pocket between them, achieving optimal density
even if eventually taken up by a normally uncooperative driver.

This mechanism also explains why the relative advantage of painting lines
accrues primarily for large values of $\alpha$: enough lines must be hit
to force such tight pockets.
Unit-length pockets are also occasionally created for other painting patterns
(i.e. $k \neq 2$),
but the probability of creating them proves too
small to compete with the benchmark strategy.

Larger gaps between painted lines necessarily implies fewer
lines available to hit.
Once these are exhausted, the ``good'' drivers must shift to an alternate
strategy.
Figure \ref{fig:density_vs_size} contrasts the densities achieved with kiss-the-bumper as a secondary
strategy (left) vs. random/R\'{e}nyi (right).
The interactions between R\'{e}nyi and painted lines can be disastrous: worse than no lines at all for $k=3$.
There the good drivers tend create pockets of two spaces, one of which becomes permanently lost to a R\'{e}nyi
driver.

In general, much of the increase in parking density from line painting comes from kiss-the-bumper as the
secondary strategy.
The painted lines primarily help by creating unit-length spaces.
Let $f(x)$ denote the number of cars a open space of length $x$ eventually contains. 
After a bumper-kiss event, $f(x) = 1 + f(x-1)$.
But a good driver hitting the line at position $y$,
yields $f(x) = 1 + f(y) + f(x-y-1)$, which is similar to the effect of a
R\'{e}nyi driver.

\section{First-Available Space Strategies}

The previous section posits drivers who exhibit no preferences
when faced with the Nirvana of multiple open spots on the street.
This utility function accurately models the behavior in the shift from
double-parking to clean streets, since the double-parked cars are uniformly distributed along the street.
It also seems relevant to the case where drivers have random destinations on the street after they leave
their vehicle.

However, in many situations drivers are likely to pounce on the first available space they see as they enter
the street.
With the new model, normal drivers will pick a random position in the first open space.
The driver's behavior is now limited to the first open pocket (i.e. the pocket closest to $x=0$) of sufficient size to
hold their unit-length vehicle.
We again assume that a $(1-\alpha)$ fraction of drivers employ the R\'{e}nyi strategy,
while the more civic-minded $\alpha$ fraction pursue one of three beneficial approaches:

\begin{itemize}

\item {\em Kiss the Bumper} --
Such drivers park directly behind the top car delimiting the first open pocket.

\item {\em Hit the Line} --
Such drivers park on the first line they see in the first open pocket, if one exists.
Otherwise, they employ kiss-the-bumper.

\item {\em Crave the Line} --
These drivers go the extra
length to park at the first line they see, even if it's not in the first open pocket.
They kiss-the-bumper if they don't find any unoccupied lines.
Intuitively, a sequence of $k$ crave-the-line drivers will create a sequence of $k-1$ open spaces of unit-length,
which when occupied by arbitrary drivers provides optimal density
over the interval of length $2k-1$.
\end{itemize}

\begin{figure}
  \centering
\subfloat[Hit-the-line vs. Kiss-the-bumper]{\includegraphics[scale=0.5]{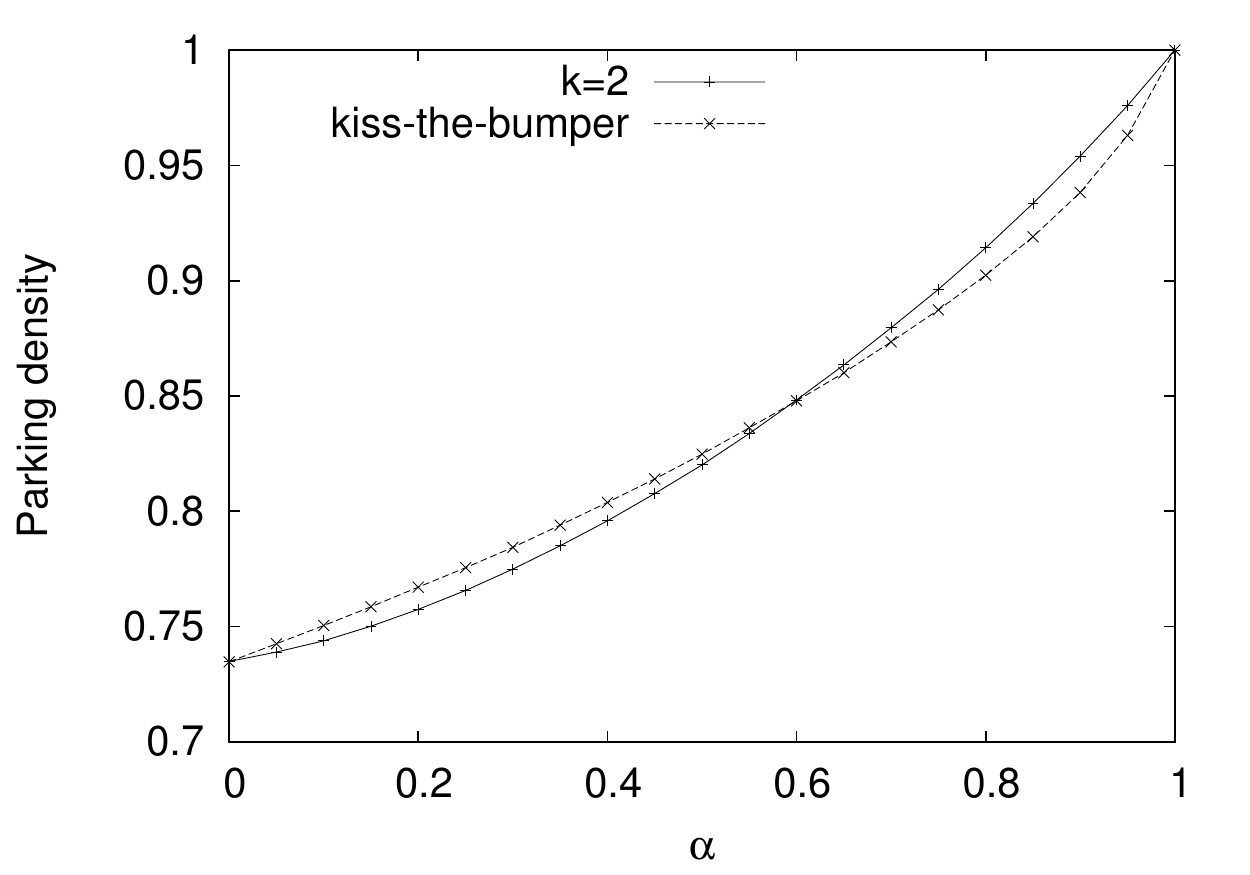}}        
\subfloat[Crave-the-line vs. Hit-the-line]{\includegraphics[scale=0.5]{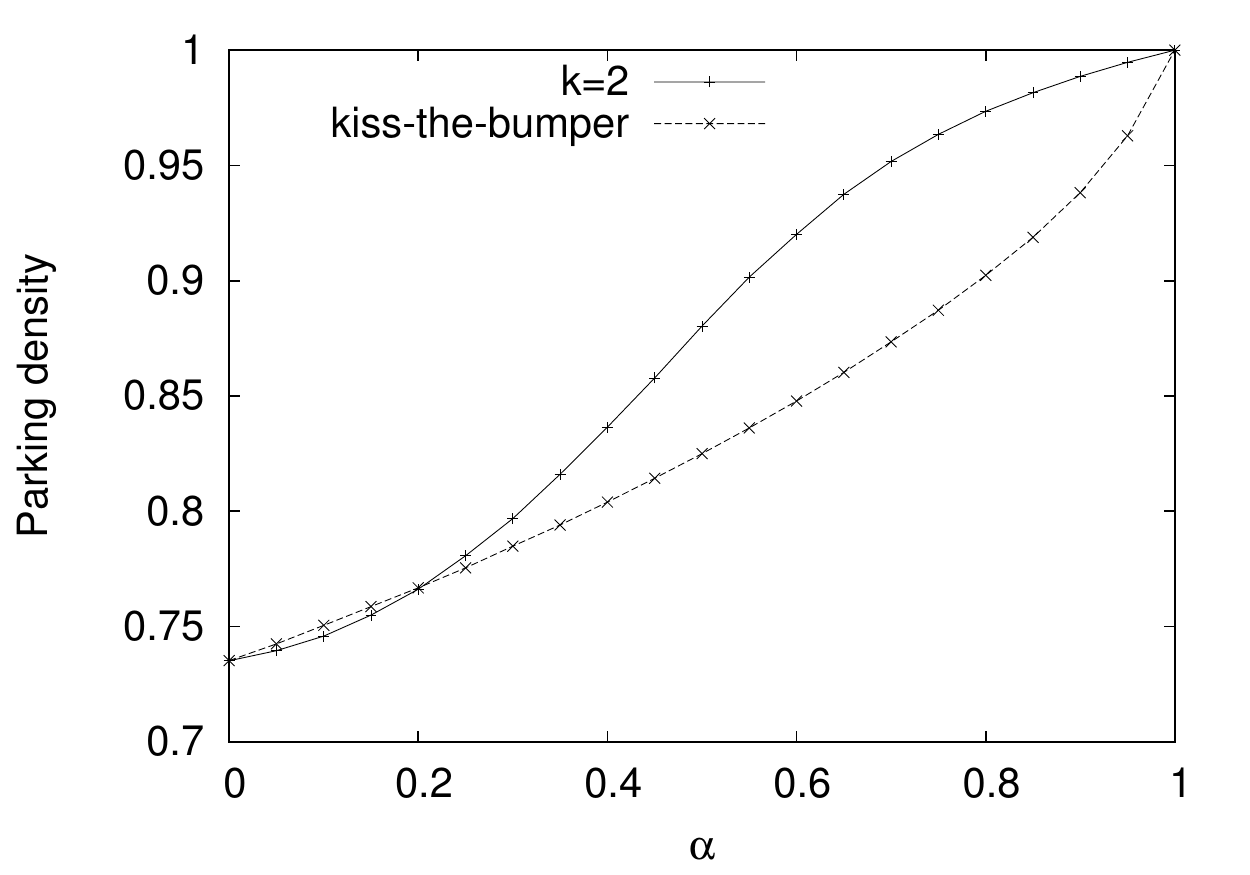}}
  \caption{Parking densities for hit-the-line, crave-the-line, and kiss-the bumper restricted to the first-available pocket.}
  \label{fig:first-pocket}
\end{figure}

Our results are shown in Figure \ref{fig:first-pocket}(a).
This figure, comparing hit-the-line to kiss-the-bumper, looks very similar to Figure \ref{main-result-figure}.
Thus driver preferences for first pocket instead of random pockets have no significant implication on the
design of line spacing or behavioral interventions.

More surprising is the excellent performance of crave the line, shown in Figure \ref{fig:first-pocket}(b).
The increase in parking density grows rapidly once one achieves a compliance rate of roughly $\alpha=0.25$.
Although we are somewhat skeptical that large numbers of drivers can be convinced to pass an open spot to hit
the next, the results do speak of the value of emphasizing proper line markings.

\section{Variable-Sized Cars}

The results reported thus far all assume unit-length vehicles, thus ignoring the natural diversity of car sizes from
compacts to SUVs.
But this non-uniform size distribution is an important factor governing urban street parking dynamics.
Certainly smaller cars have an easier time finding a parking in the city than big ones,
which goes a long way towards explaining the popularity of half-size Smart cars in urban environments.

Variable car lengths require us to consider whether our optimization criteria
is to maximize the number of cars resting on the street, or the fraction of the total street length
being utilized by parked cars.
The former could best be done by only allowing half-sized cars to park, while the later would reserve a spot
for the largest car which could fit in it.
For the purposes of this paper, we seek to maximize the number of parked vehicles without explicitly favoring
small cars over large ones.
Thus cars are served at random as per their length-frequency distribution, but the process continues until no more empty spots for minimum-length cars remain.

In this section, we study two different models of non-uniform car lengths.
Section \ref{uniform-distribution} considers a model where cars are uniformly distributed over a size range.
Section \ref{real-world-distribution} uses car sales data and representative Manhattan streets to strive for
more accurate modeling.

\subsection{Uniform Distribution of Lengths}
\label{uniform-distribution}

To capture the variance of car lengths, we consider car lengths as generated by a uniform distribution in the range
$[1-d,1+d]$, for some value of $d$.
The previously studied case of unit-length cars corresponds to $d=0$.

\begin{figure}
  \centering
\subfloat[Any pocket]{\includegraphics[scale=0.43]{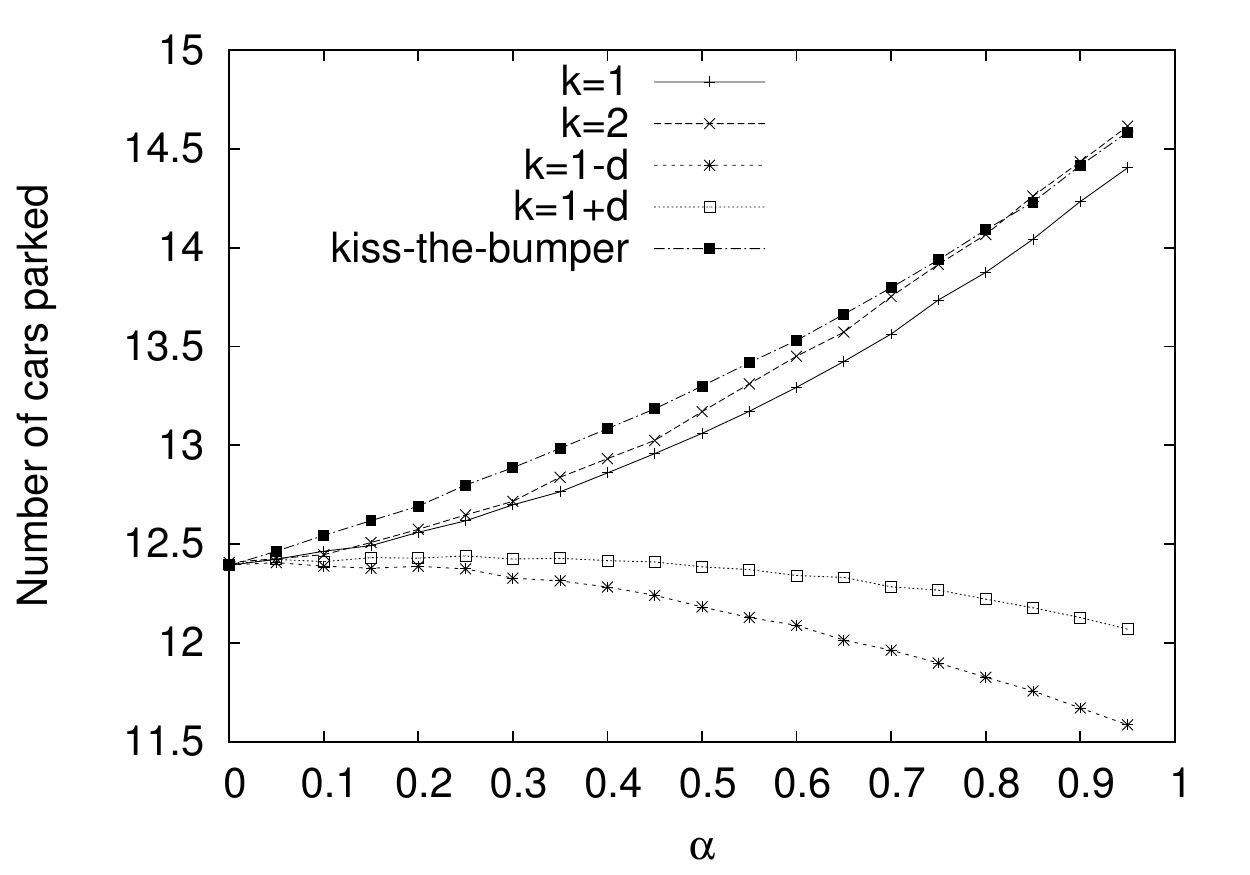}}
\subfloat[First pocket]{\includegraphics[scale=0.43]{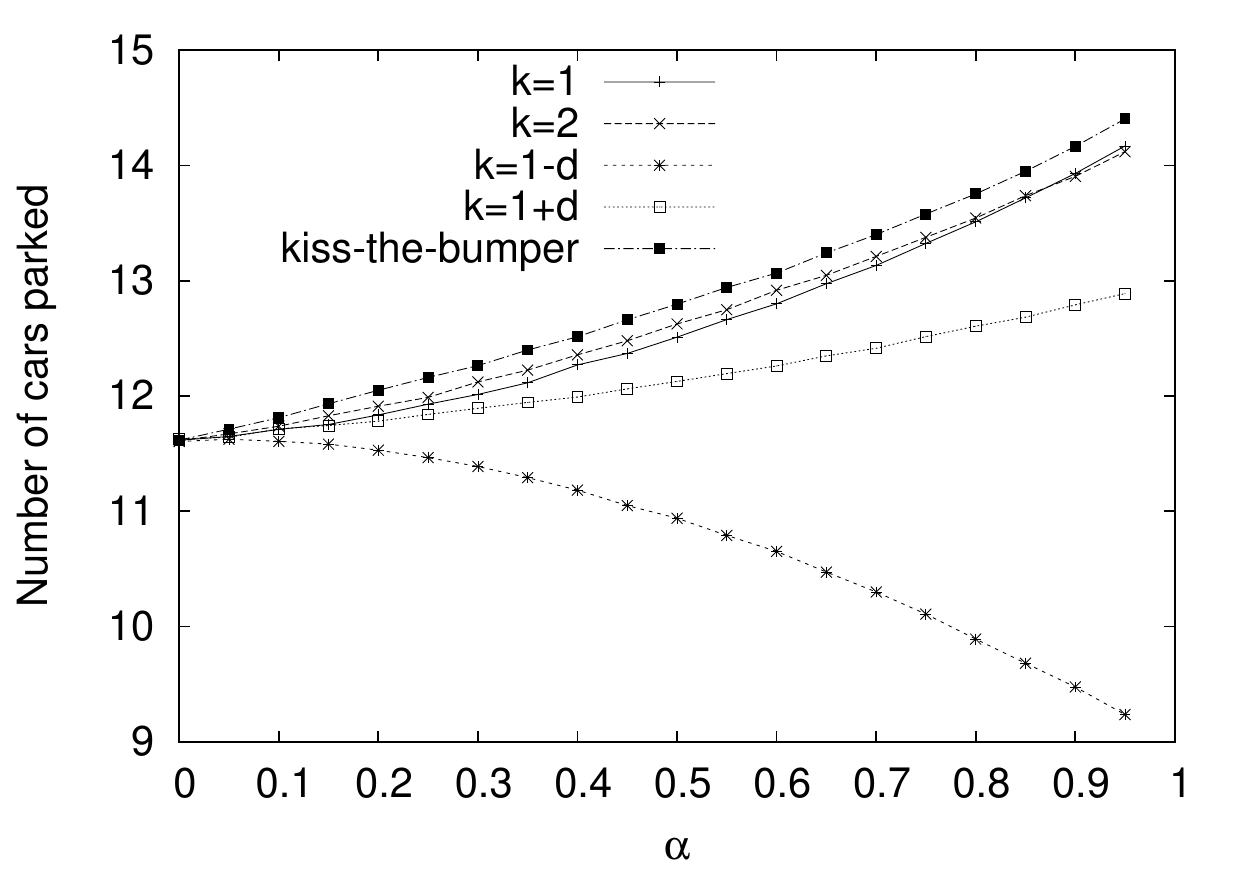}}
\subfloat[Crave the line]{\includegraphics[scale=0.43]{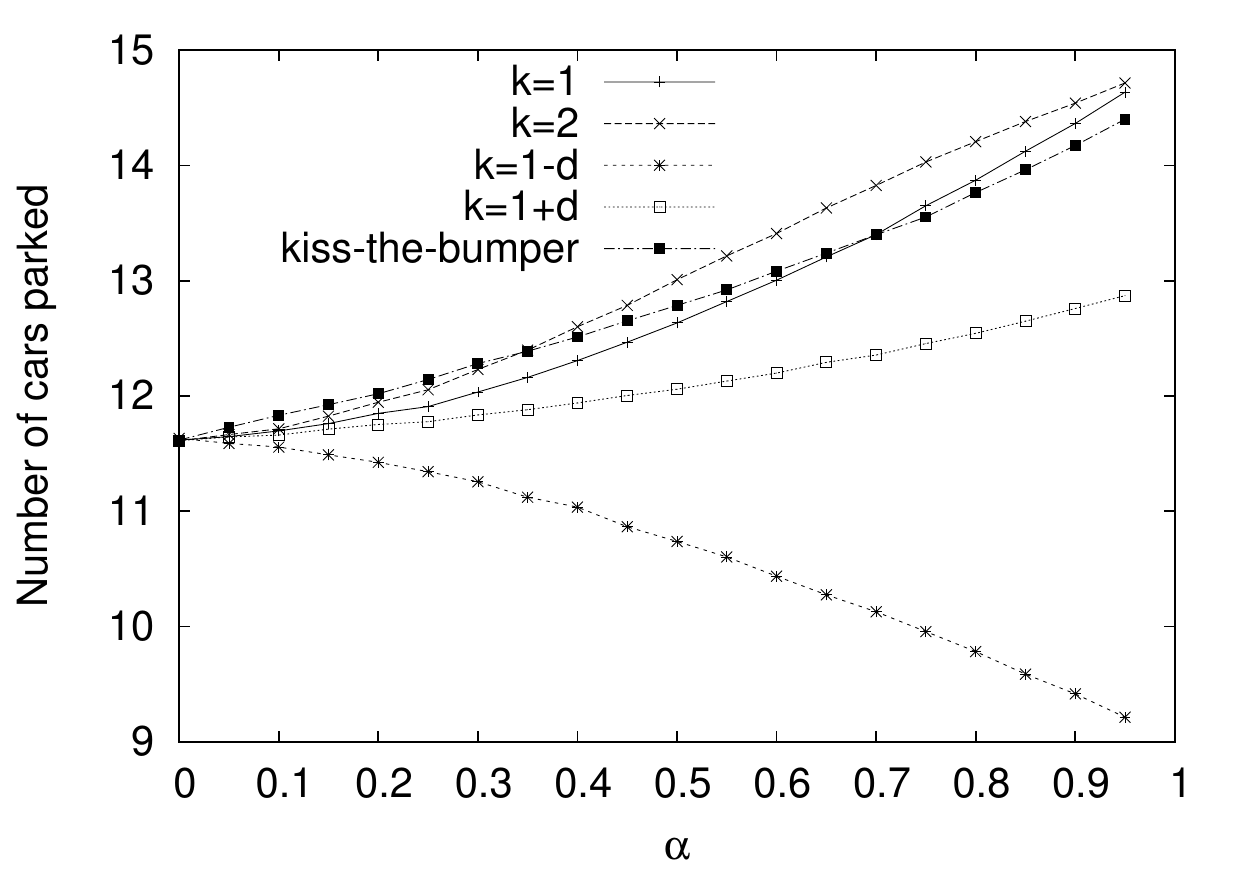}}
  \caption{Parking densities for car lengths drawn uniformly from [0.75,1.25].}
  \label{fig:diffsizesimresult}
\end{figure}

We evaluated four gap sizes for the line markings for each of the three pocket-selection models
previously described (random pocket, first pocket, and crave-the-line).
The gap sizes we study are $1-d$, $1+d$, $1$ or $2$.

The results for $d=0.25$ are shown in Figure \ref{fig:diffsizesimresult}, with
the results for $d=0.125$ and $d=0.5$ quite similar.
As expected, the $k = 2$ gap length generally outperforms all other line markings, although there are small windows where other spacings dominate
for high $\alpha$ in hitting or craving the line.
Making the gap larger or smaller than an integral size of the car did not perform well.
Indeed the tight gap spacing of $1-d$ results in strictly fewer parked cars as compliance increases.

\subsection{Real World Distributions}
\label{real-world-distribution}

\begin{figure}
\centering
   \includegraphics[scale=0.60]{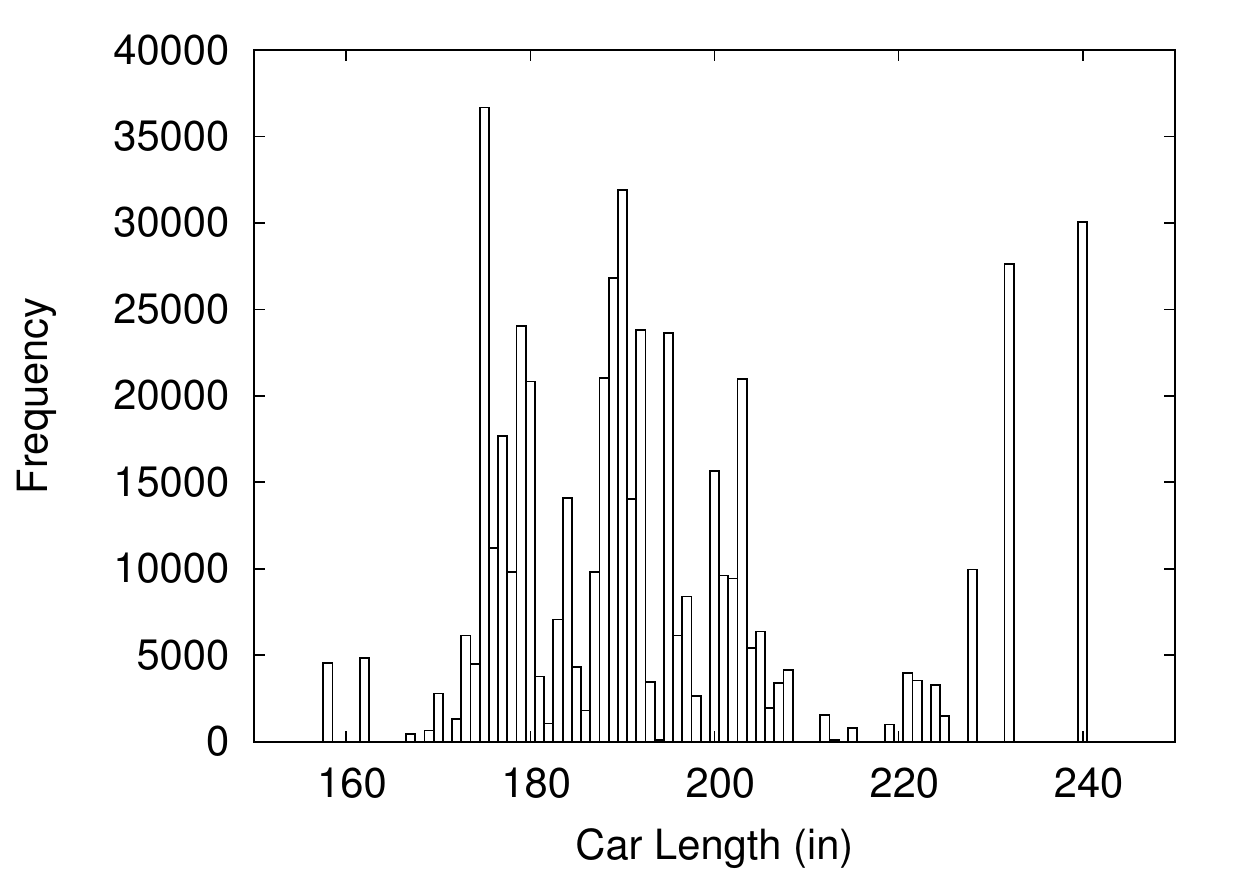}
\caption{North American car sales in 2010 as a function of body length.
\label{length-histogram}}
\end{figure}

The optimal line spacing depends on the distribution of the car lengths
driven by local residents.
To better capture the dynamics of real world parking,
we used new car sales figures as a proxy for the length distribution in
use on today's roads.
In particular, we compiled the January 2010 sales for each current model from
GM, Ford, Honda, Toyota, Nissan, Chrysler, and Kia.
These seven companies together control 85\% of the automobile market share in North America. 
Figure \ref{length-histogram} presents the relative frequencies of cars by body length.
It is strikingly irregular, and not particularly well approximated by the uniform distribution
employed in the previous section.

We performed simulation experiments on two typical sizes of Manhattan streets.
The Broadway-to-Amsterdam and Amsterdam-to-Columbus blocks of West 92nd Street
measure in at 340 feet and 831 feet, respectively.
We assume that adjacent parked cars must be at least one foot (12 inches)
apart from each other.

\begin{figure}
  \centering
\subfloat[Short block]{\includegraphics[scale=0.5]{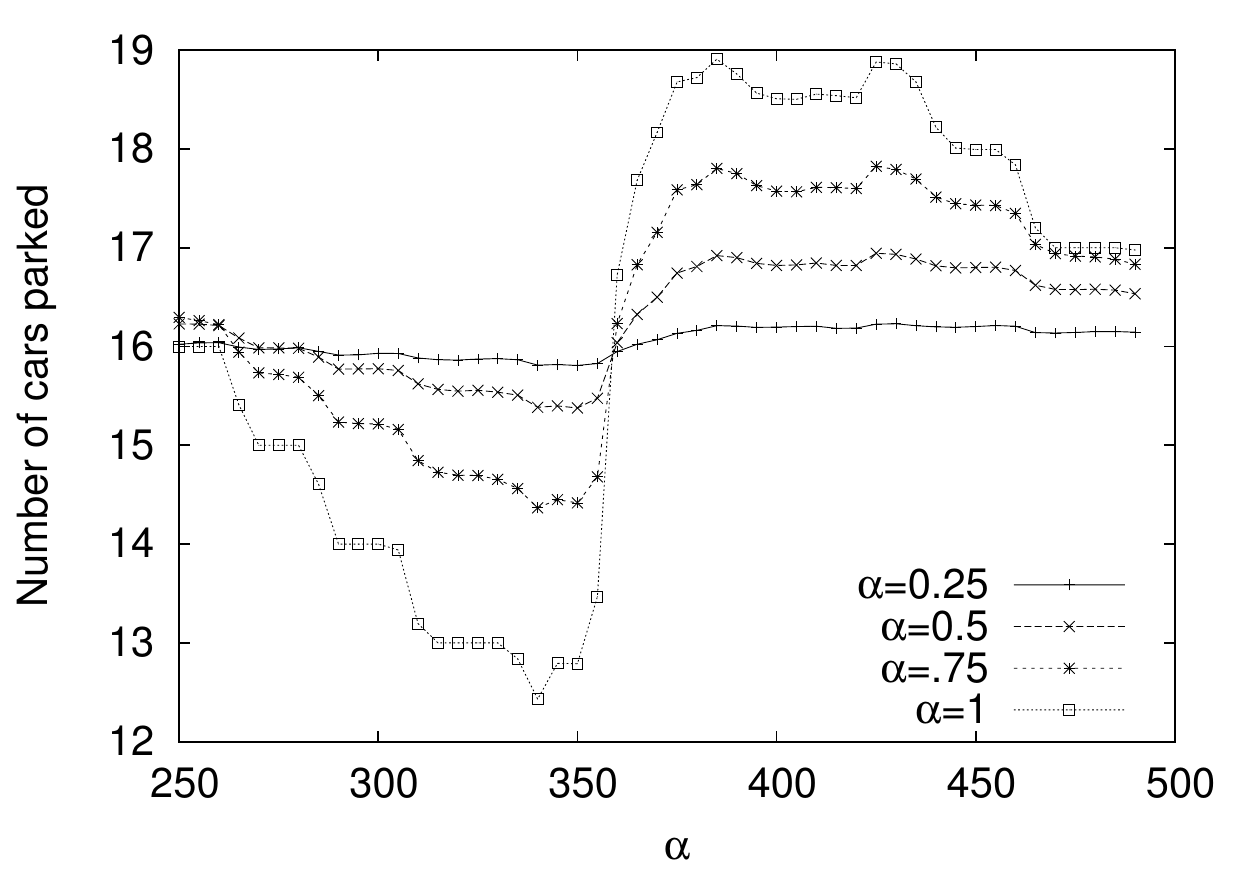}}
\subfloat[Long block]{\includegraphics[scale=0.5]{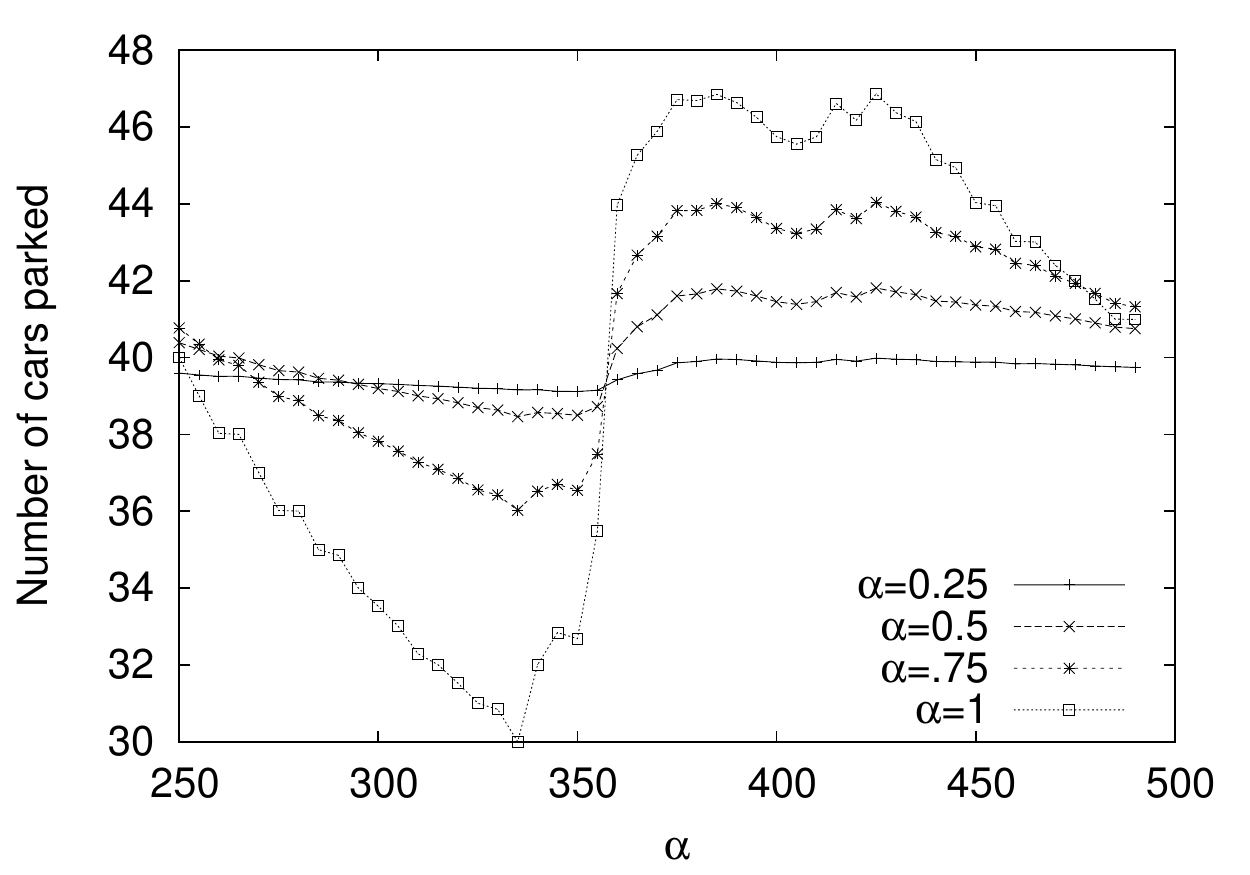}}
  \caption{Simulation result for real world data: the number of parked cars as a function of the distance between painted lines.}
  \label{fig:real_density_vs_size}
\end{figure}

Figure \ref{fig:real_density_vs_size} presents our results on the expected number of cars parked on both short and long blocks as a function
of the gap between painted lines for four values of $\alpha$.
The optimal gap between painted lines was determined to be 385/390 inches on short/long blocks,  respectively.
This density oscillates depending upon how this gap compares to the length of a typical car.
As previously observed, optimal density is achieved when the gap is roughly twice the average car length.
Making the gap too short relative to this length significantly degrade performance, well below that of unmarked streets.
The magnitude of this oscillatory behavior increases strictly with $\alpha$.

\begin{figure}
  \centering
\subfloat[Short block]{\includegraphics[scale=0.5]{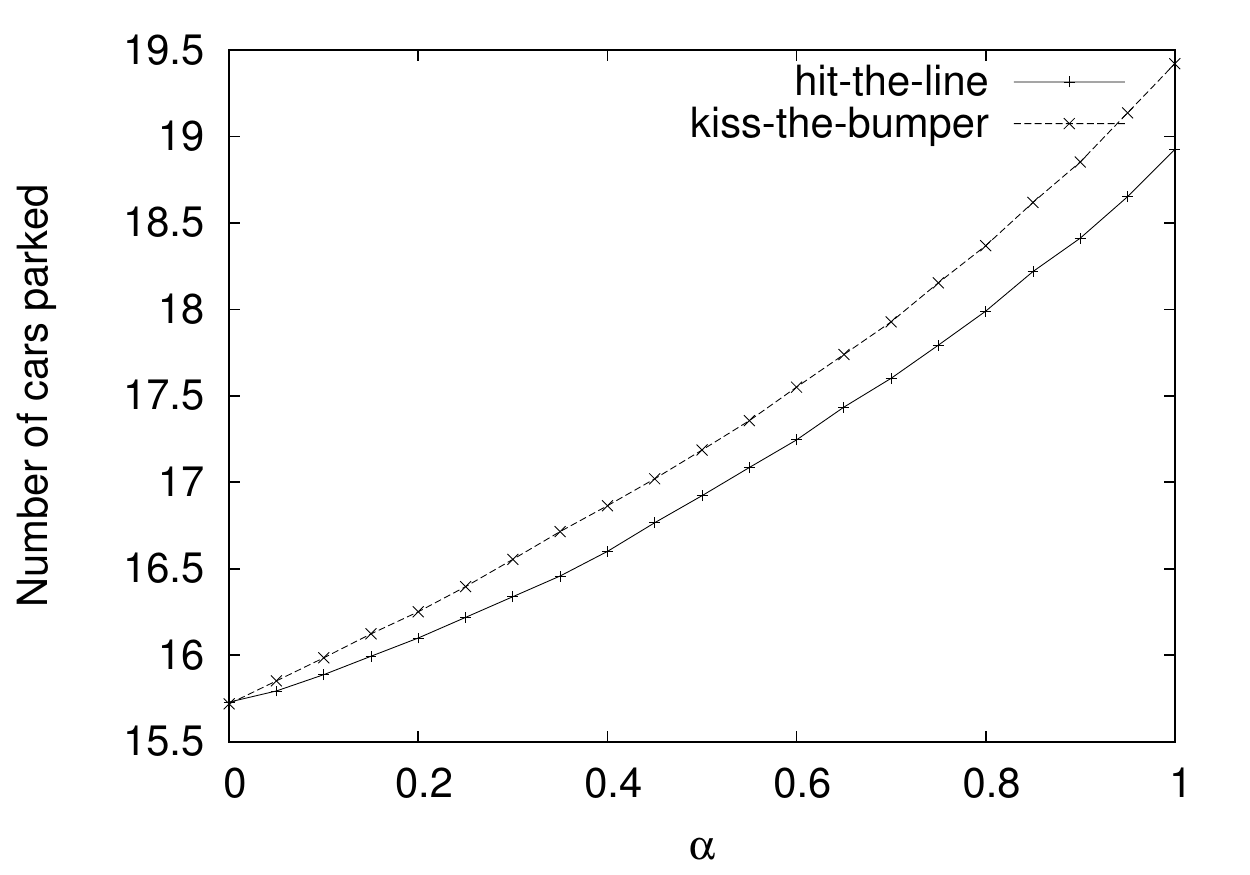}}
\subfloat[Long block]{\includegraphics[scale=0.5]{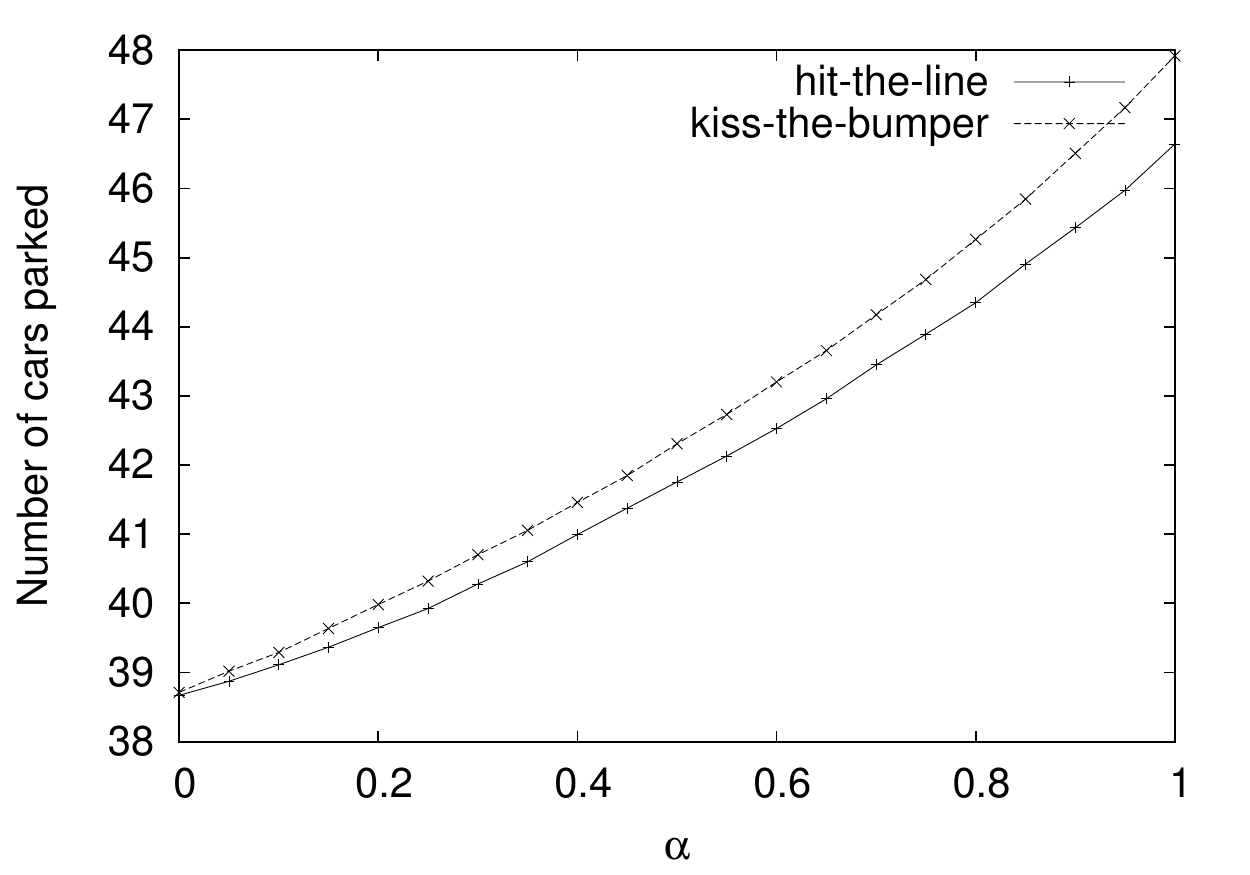}}
  \caption{Simulation result for real world data: the number of parked cars as a function of the compliance rate $\alpha$.}
  \label{fig:real_density_vs_alpha}
\end{figure}

Figure \ref{fig:real_density_vs_alpha} gives an alternate view of this data, showing the average number of cars parked on both short and long
blocks as a function of $\alpha$ for optimally-sized gaps.
The number of cars increases continually with greater compliance, with greater improvement per $\Delta \alpha$ as $\alpha$ increases.

\section{Conclusions}
\label{conclusions}
We have studied the impact of two behavioral interventions (encouraging kiss-the-bumper or hit-the-line parking driving) to improve space
utilization in city streets.
Generally kiss-the-bumper provides better utilization for a given level of compliance, however the absence of visual cues makes it difficult and
expensive to alter current habits.
If painting guide lines can increase compliance ($\alpha$) by $0.1$, 
then painting guide lines two car-widths apart is the right strategy.

With the proper street markings, one can generally squeeze one extra car into the shorter street on average at $\alpha = 0.5$, and the same on the
longer street at the lower compliance of $\alpha = 0.4$. 
As Manhattan is roughly a grid of 200 streets by 15 avenues, each consisting of two sides, these results argue that the parking capacity of the city could
be increase by roughly 6,000 vehicles in steady-state by employing the right line painting strategy.
We 
hope that our results can lead to a pilot painting study to see how well theory matches practice.

\section{Acknowledgments}
We wish to thank the Algorithms Reading Group at Stony Brook
for helpful discussions.

\bibliography{refs}

\begin{thebibliography}{10}

\bibitem{NPA-10}
National~Parking Association.
\newblock The dimensions of parking, 2010.

\bibitem{BS-70}
B.~Blaisdell and H.~Solomon.
\newblock On random sequential packing in the plane and a conjecture of
  palasti.
\newblock {\em J. Appl. Prob.}, 7:667--698, 1970.

\bibitem{cmp}
A.~Cadilhe, N.~Araujo, and V.~Privman.
\newblock Random sequential adsorption: from continuum to lattice and
  pre-patterned substrates.
\newblock {\em J. of Physics: Condensed Matter}, 19, 2007.

\bibitem{CK-98}
R.~Cassady and J.~Kobza.
\newblock A probabilistic approach to evaluate strategies for selecting a
  parking space.
\newblock {\em Transportation Science}, 32:30--42, 1998.

\bibitem{CSBIM-01}
A.~Chrest, M.~Smith, S.~Bhuyan, M.~Iqbal, and D.~Monahan.
\newblock {\em Parking Structures: Planning, Design, Construction, Maintenance,
  and Repair}.
\newblock Kluwer Academic, Boston MA, 2001.

\bibitem{bowlinggreen}
Kentucky City~of Bowling~Green.
\newblock On-street parking.
\newblock http://www.bgky.org/publicworks/planningdesign/
  transportation/pdf/On-Street-Parking.pdf, 2002.

\bibitem{ClaySimanyi-14}
M.~P. {Clay} and N.~J. {Simanyi}.
\newblock {Renyi's Parking Problem Revisited}.
\newblock {\em ArXiv e-prints}, June 2014.

\bibitem{onparking}
A.~Dvoretzky and H.~Robbins.
\newblock On the parking problem.
\newblock {\em Publ. Math. Inst. Hung. Acad. Sci.}, 9:209--224, 1964.

\bibitem{finch}
S.~Finch.
\newblock {\em R\'{e}nyi's Parking Constant}, pages 278--284.
\newblock Cambridge University Press, London, 2003.

\bibitem{KPP-04}
H.~Kellerer, U.~Pferschy, and P.~Pisinger.
\newblock {\em Knapsack Problems}.
\newblock Springer, 2004.

\bibitem{LMM-02}
A.~Lodi, S.~Martello, and M.~Monaci.
\newblock Two-dimensional packing problems: A survey.
\newblock {\em European J. Operations Research}, 141:241--252, 2002.

\bibitem{Mannion-64}
D.~Mannion.
\newblock Random space-filling in one dimension.
\newblock {\em Publ. Math. Inst. Hung. Acad. Sci.}, 9:143--154, 1964.

\bibitem{ADA-10}
United States~Department of~Justice.
\newblock 2010 standards for accessible design.
\newblock Print, 2010.

\bibitem{Palasti-60}
I.~Palasti.
\newblock On some random space filling problems.
\newblock {\em Publ. Math. Inst. Hung. Acad. Sci.}, 5:353--359, 1960.

\bibitem{Renyi-58}
A.~R\'{e}nyi.
\newblock On a one-dimensional problem concerning random space-filling.
\newblock {\em Publ. Math. Inst. Hung. Acad. Sci.}, 3:109--127, 1958.

\bibitem{Rojas-06}
D.~Rojas.
\newblock Revenue management techniques applied to the parking industry.
\newblock M.S. Thesis, University of South Florida, 2006.

\bibitem{Tepper}
C.~Trillin.
\newblock {\em Tepper Isn't Going Out}.
\newblock Random House, 2001.

\bibitem{Routledge-04}
K.~Vincente.
\newblock {\em The Human Factor}.
\newblock Routledge, 2004.

\end{thebibliography}
\end{document}